\documentclass[aps,prd,showkeys,showpacs,amssymb,
amsfonts,epsf,preprintnumbers,nofootinbib,superscriptaddress]{revtex4}

\usepackage[dvips]{graphicx}
\usepackage{bm,latexsym,amsmath,amssymb,amsfonts,color}

\newcommand\beq{\begin{eqnarray}}
\newcommand\eeq{\end{eqnarray}}

\newcommand{\nn}{\nonumber}

\begin{document}

\title{Origin of the universe: A hint from Eddington-inspired Born-Infeld gravity}

\author{Hyeong-Chan Kim}
\email{hckim@ut.ac.kr}
\affiliation{School of Liberal Arts and Sciences, Korea National University of Transportation, Chungju 380-702, Korea}

\begin{abstract}
We study the `initial state' of an anisotropic universe in Eddington-inspired Born-Infeld gravity filled with a scalar field, whose potential has various forms.
With this purpose, the evolution of a spatially-flat, homogeneous anisotropic Kasner universe is studied.
We find an exact evolution of universe for each scalar potential by imposing a maximal pressure condition.
The solution is shown to describe the initial state of the universe.
The state is regular if the scalar potential increases not faster than the quadratic power for large field values.
We also show that the anisotropy does not raise any defect in early universe contrary to the case of general relativity.
\end{abstract}
\maketitle

\section{Introduction}
What is the origin of the universe?
This is one of the oldest quest of human.
After the discovery of the Hubble's law in 1929, people started to believe that the universe originated from big bang.
However, the big-bang theory was soon confronted with obstacles such as the flatness, horizon, and monopole problems.
Most of them was resolved by introducing inflation theory in 1980.
However, we still confront many things to discover on the origin of the universe, such as the origin and physics before and at the early stage of the inflation.
In addition, even the inflation theory connotes problems such as fine-tuning~\cite{Boyle:2005ug} and low entropy initial state~\cite{1983Natur.304...39P}.
In the framework of the Einstein's general relativity (GR), the question on the initial state of the universe is hard to answer.
This is because GR actually predicts that a singularity appears at the beginning of the universe.

There are a few proposals which avoid initial singularities.
An interesting proposal is to assume that the universe begins with no origin through eternal inflation~\cite{Carroll:2005it}.
However, the inflation requires extremely special initial conditions.
One may consider the universe has a quantum cosmological origin~\cite{Vilenkin,HH}.
In fact, this suggestion naturally give rises to the necessity of quantum gravity giving quantum corrections to GR.
One may expect that, with proper consideration of quantum gravitational effects, a gravitational theory must be free from singularity.
In this work, we show that the Eddington-inspired Born-Infeld (EiBI) theory of gravity, recently proposed by Ban\~{a}dos and Ferreira~\cite{Banados:2010ix},
provides an opportunity to analyze the initial state of the universe.

The EiBI theory of gravity is described by the action \begin{eqnarray}\label{maction}
S_{{\rm EiBI}}=\frac{1}{\kappa}\int
d^4x\Big[~\sqrt{-|g_{\mu\nu}+\kappa
R_{\mu\nu}(\Gamma)|}-\lambda\sqrt{-|g_{\mu\nu}|}~\Big]+S_M(g,\Phi),
\end{eqnarray}
where $|g_{\mu\nu}|$ denotes the determinant of $g_{\mu\nu}$,
$\lambda$ is a dimensionless parameter which is related with the cosmological constant, and $\kappa$ is an additional parameter.
Throughout this paper, we use the reduced Planck units, $c=1=8\pi G$.
In this theory the metric $g_{\mu\nu}$ and the connection $\Gamma_{\mu\nu}^{\rho}$
are treated as independent fields.
The Ricci tensor $R_{\mu\nu}(\Gamma)$ is evaluated solely from the connection,
and the matter field $\Phi$ couples not with $q_{\mu\nu}$ but with $g_{\mu\nu}$.
In Refs.~\cite{Pani:2011mg,Pani:2012qb}, the authors considered the modification of Poisson equation in EiBI gravity, and obtained singularity-free solutions for the compact stars composed of pressureless dust and polytropic fluids.
In Refs.~\cite{DeFelice:2012hq,Avelino:2012ge,Avelino:2012qe,Casanellas:2011kf,EscamillaRivera:2012vz,Avelino:2012ue,Liu:2012rc,Delsate:2012ky}, the cosmological and astrophysical aspects on the EiBI theory were studied including solar system test showing that it is compatible with all current observations.
However, recently, it was also announced that curvature singularities happen at the surface of polytropic stars~\cite{Pani:2012qd}.
The initial state leading to a chaotic inflation was studied in Ref.~\cite{Cho:2013pea}.
Stars in the EiBI gravity were also been studied in Refs.~\cite{Yang:2013hsa,Sham:2013sya,Harko:2013wka,Sham:2012qi} to show that the neutron and quark stars are more massive than their GR counterparts.

It is important to check the singularity free nature because it is one of the main motivations of the EiBI theory.
However, a perfect fluid analogy of fundamental particles is not always successful in the extreme situations such as the early universe.
To investigate the non-singular property seriously one needs to deal the universe filled with fields whose equation of states vary depending on physical situations.
With this in mind, we consider a universe filled with a real scalar field with action
\begin{equation}
S_M = \int d^4 x \sqrt{|g|} \left[-\frac12 g^{\mu\nu}(\partial_\mu\phi)(\partial_\nu \phi) -V(\phi)\right].
\end{equation}
The equation of state parameter of the scalar field in homogeneous space-time depends on the relative ratio between the kinetic energy and the potential energy,
$w=(\dot \phi^2/2V-1)/(\dot\phi^2/2V+1)$.
If the ratio is large, $w\to 1$ and if it goes to zero, $w\to -1$.
For a given scalar potential, we cannot determine the initial state of the universe from the experience of perfect fluid because the equation of state varies with time.

\section{Anisotropic universe with a scalar field}
After the discovery of Planck observations~\cite{Ade:2013ktc} that nontrivial axial anomalies exist in the cosmic microwave background radiation, the anisotropic universe forms a new trend in cosmology and astrophysics.
The metric for a spatially flat homogeneous space-time is
\begin{eqnarray}
g_{\mu\nu} dx^\mu dx^\nu &=& -dt^2 + a^2(t) \left[e^{2(\beta_++\sqrt{3}\beta_-)}dx^2 +e^{2(\beta_+-\sqrt{3}\beta_-)}dy^2 + e^{-4\beta_+} dz^2\right] ,
\end{eqnarray}
where $a(t)$ and $\beta_\pm(t)$ denote the scale factor and the anisotropies, respectively.
Most previous works on EiBI gravity were based on a perfect fluid.
Most interestingly, the universe driven by radiation is free from the initial singularity~\cite{Banados:2010ix};
the universe experiences a bouncing with a finite size for $\kappa<0$, or there is a state of minimum size for which one takes infinite time to reach from the present for $\kappa>0$.
The latter is interpreted as the ``nonsingular initial state" of the universe.
The ``nonsingular initial state" is also present if the equation of state parameter, $w=p/\rho$, of a fluid is positive~\cite{Cho:2012vg}.
Another nonsingular initial state was known to exist as a de Sitter space-time for the dust-filled universe with $w=0$~\cite{Cho:2012vg} and in the presence of a massive scalar field~\cite{Cho:2013pea}.

The Hubble parameter as an equation for gravity+scalar field in isotropic space-time was first obtained in Eq.~(46) in Ref.~\cite{Scargill:2012kg} with $\lambda=1$.
There, it was also shown that GR is recovered to leading order in later times of expanding universe.
In the case of an anisotropic universe of Kasner type, the Hubble parameter is modified to be
\begin{eqnarray} \label{H}
 H &=&\frac{1}
    {(\lambdabar + V)^2 +\frac12 \dot \phi^4 } \left\{ -\frac{1}{2}\left(\lambdabar+V + \frac12\dot \phi^2\right)
     V'(\phi)\dot \phi\right.  \\
 &\pm &\left.\frac{1}{\sqrt{3}}\Big(\lambdabar+V-\frac{\dot \phi^2}2\Big) \, \left[\Big(\lambdabar+V + \frac12\dot \phi^2\Big)^{3/2}
    \Big(\lambdabar +V - \frac12\dot \phi^2\Big)^{3/2}+
    \frac1{\kappa}\Big(\lambdabar+V + \frac12\dot \phi^2\Big)\Big( \dot \phi^2 - V-\lambdabar\Big) + \frac{6c^2}{ a^6}
     \right]^{1/2}\right\} , \nn
\end{eqnarray}
where $\lambdabar=\lambda/\kappa$, $c^2=c_+^2+c_-^2$ and we are interested in the case with $\kappa, \lambda >0$.
The isotropic space version of this equation was given in Ref.~\cite{Cho:2013usa}.
The anisotropy satisfies
\begin{equation}\label{beta}
\dot \beta_{\pm} = \frac{c_\pm}{(\lambdabar+V+\frac{\dot \phi^2}2) a^3} ,
\end{equation}
which was obtained exactly the same way as that in Ref.~\cite{Cho:2012vg}.
The scalar field equation is given by
\begin{equation} \label{eom:phi}
\ddot \phi +3 H \dot \phi + V'(\phi) =0.
\end{equation}

\subsection{Constant potential case}
\begin{figure}[btph]
\begin{center}
\begin{tabular}{ll}
\includegraphics[width=.5\linewidth,origin=tl]{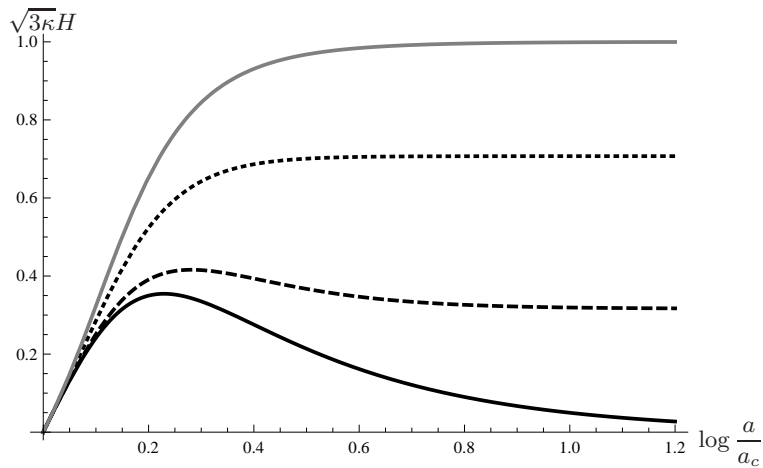}
\end{tabular}
\put (-260,85) {$\sqrt{3\kappa}H $}
\put (-2,-73) { $\displaystyle \log\frac{a}{a_c} $}
\end{center}
\caption{The Hubble parameter with respect to the scale factor for various cosmological constant.
Here, the cosmological constant is given by $\Lambda=0$, $0.1/\kappa$, $0.5/\kappa$, and $1/\kappa$, respectively from the bottom.
} \label{fig:Hconst}
\end{figure}
It was known that the EiBI theory in vacuum is equivalent to GR~\cite{Banados:2010ix}.
In the case of a constant potential, $V(\phi)=V_0$, the potential plays the same role as a cosmological constant with $\Lambda\equiv V_0+\lambdabar-1/\kappa$ if one ignore the scalar kinetic energy.
Even with scalar dynamics, one can find an exact solution, which resembles the ``nonsingular initial state" in the perfect fluid case.
Eq.~\eqref{eom:phi} is integrated to give
$
\dot \phi = \sqrt{2\left(1/\kappa+\Lambda\right)} (a_c/a)^3,
$
where $a_c$ is a critical scale whose physical implication will be clear soon.
The Hubble parameter becomes
\begin{equation}\label{H:const}
 H(a) = \frac{1}{\sqrt{3\kappa}} \frac{1-\frac{a_c^6}{a^6}}
    {1 +\frac{2 a_c^{12}}{a^{12}} }
\left[ \left(1+\kappa \Lambda\right)
    \left(1-\frac{ a_c^{12}}{a^{12}}\right)^{3/2}-1+\left(1+\frac{6\kappa c^2}{a_c^6} \right)\frac{a_c^6}{a^6}+\frac{2a_c^{12}}{a^{12}} \right]^{1/2},
\end{equation}
where we take the positive sign in Eq.~\eqref{H}, because we want to describe an expanding universe.
The Hubble parameter is plotted in Fig.~\ref{fig:Hconst}.
The terms inside the square braket can be expanded as, for large $a$,
$$
[\cdots] =\kappa \Lambda +\left( 1+\frac{6\kappa c^2}{a_c^6} \right)\frac{a_c^6}{a^6}+\frac12\left(1-3\kappa \Lambda\right) \frac{a_c^{12}}{a^{12}} +\cdots
$$
The Hubble parameter goes to zero at the minimum value of the scale factor,
$
a= a_c.
$
Around the value, the scale factor behaves as
\begin{equation} \label{a1:t}
a(t) = a_c + \delta_0 e^{2 \sqrt{\frac{6}{\kappa}} t} ,
\end{equation}
which presents the same ``nonsingular initial state" as the perfect-fluid counterpart.

The scalar velocity with respect to the scale factor can also be written as
\begin{equation} \label{phi:aconst}
\frac{d\phi}{da} = H^{-1}\frac{d\phi}{dt}
  =\sqrt{\frac{2\left(1/\kappa+\Lambda\right)}{3\kappa}} \left(\frac{a_c}{a}\right)^3
     \frac{1-\frac{a_c^6}{a^6}}
    {1 +\frac{2 a_c^{12}}{a^{12}} }
\left[ \left(1+\kappa \Lambda\right)
    \left(1-\frac{ a_c^{12}}{a^{12}}\right)^{3/2}-1+\left(1+\frac{6\kappa c^2}{a_c^6} \right)\frac{a_c^6}{a^6}+\frac{2a_c^{12}}{a^{12}} \right]^{1/2}.
\end{equation}
Eqs.~\eqref{H:const} and \eqref{phi:aconst} can be regarded as an approximate solution for cases when the scalar potential approaches a constant value for large $\phi$.

\section{Initial state of the universe}
From the experience of the perfect fluid in Ref.~\cite{Cho:2012vg}, the initial state of a universe appears to depend on the equation of state of its matter contents for positive $\kappa$.
We inspect whether this dependence is extendable to the case of a scalar field or not.
In Ref.~\cite{Cho:2013pea}, the authors have shown that a {\it maximal pressure condition},
\begin{equation}\label{cond}
\frac12 \dot \phi^2 - V(\phi) -\lambdabar = 0 \quad  \Longrightarrow \quad \frac{d\phi}{dt} = U(\phi),
\quad \mbox{where} \quad U(\phi) \equiv\sqrt{2(V(\phi)+\lambdabar)}
\end{equation}
combined with the Hubble parameter equation
\begin{equation} \label{H:V}
H = -\frac{2}{3} U'(\phi)
\end{equation}
provides an exact solution, which was called as {\it maximal pressure solution} (MPS), of the equation of motion of the EiBI gravity.
In Eq.~\eqref{cond}, we choose the positive sign.
Solutions with $\frac{d\phi}{dt}=-U$ can be obtained by taking $t\to -t$.
The reality of $\dot \phi$ requires $V(\phi) \geq -\lambdabar$.
The early time behavior of the MPS describes the initial state of the universe~\cite{Cho:2013pea}.
Now, the MPS is given by
\begin{equation}\label{phi:t}
\mathfrak{T}(\phi)\equiv \int^\phi \frac{d\phi'}{U(\phi')}=\int^t d\tau \quad \Longrightarrow \quad
\phi(t) = \mathfrak{T}^{-1}(t).
\end{equation}
Eq.~\eqref{H:V} is integrated to give the scale factor
\begin{eqnarray} \label{a:t}
\frac{\dot a}{a} = -\frac23 U'(\phi)  = -\frac{2\dot U}{3 U} \quad \Longrightarrow \quad a(t) = a_0 \big[U(\phi)\big]^{-2/3},
\end{eqnarray}
where $a_0$ is an integration constant.
The metric function describing the anisotropy is
\begin{equation}\label{beta}
\beta_\pm = \int^t dt \frac{c_\pm}{ a_0^3 } =  \frac{c_\pm}{ a_0^3 }(t-t_\beta).
\end{equation}
The divergence of the anisotropy,
$
I_\pm = \dot \beta_\pm/H = -3 c_\pm/(2 a_0^3 U'(\phi)) ,
$
happens only when $H=0$.
However, the geometry there is nonsingular.

\subsection{The initial state}
We next write down the early time behaviors of the universe for various nonsingular potentials.
The `initial state' of an expanding universe is determined by the asymptotic form of the scalar potential for large field value.
This is simply because the size of the scalar field should be very large in the early universe.
For the initial state~\eqref{cond}, the scalar field always increase with time, $\dot \phi>0$.
Therefore, once $\dot \phi (\propto U(\phi))$ is positive at a time $t_0$, it is always positive at earlier times $t<t_0$, implying $\phi \to -\infty$ in the early universe.
Let us assume that the asymptotic form of the scalar potential takes the form,
\begin{equation} \label{U:n}
  U=  \mu |\phi|^{n+1}\Rightarrow V = \frac{\mu^2}2 \phi^{2n+2}-\lambdabar,
\end{equation}
where we choose $\mu \geq 0$ without loss of generality.
For the case of $n=-1$, the scalar potential approaches to a constant and the corresponding early time behavior was given in Eq.~\eqref{a1:t}.

For $n\neq 0, ~-1$, from Eq.~\eqref{phi:t}, the MPS behaves as
\begin{eqnarray}\label{sol:n}
\phi = -\frac1{(n\mu \,t)^{\frac1n}}, \qquad
a(t) = a_0 U^{-2/3}  = \frac{a_0}{\mu^{2/3}}
	\big(n\mu\, t\big)^{\frac{2(n+1)}{3n}}.	
\end{eqnarray}
For $n>0$,  the domain of time is $t\in (0 , \infty)$.
The scalar field evolves from $-\infty$ to $0$ with time.
The scale factor monotonically increases from $0$ to $\infty$.
It describes a power-law-inflating or a decelerating universe, respectively for $0<n< 2$ or $n>2$.
Observing backward in time, the scale factor decreases to zero in a finite time and the geometry becomes singular at $t=0$.
Therefore, in this case, the universe starts from a singularity similarly to the case of big-bang cosmology and the nonsingular initial state is absent.
The case with $n=0$ is described by an initial de Sitter universe, which was studied in Ref.~\cite{Cho:2013pea}.
For $n< 0,~ (\neq -1)$,
the domain of time is $(-\infty,0]$ and the scalar field runs from $-\infty$ to $0$.
For $-1< n<0$, the scale factor expands with accelerating rates.
For large negative $t$, both the acceleration and the velocity of the scale factor go to zero.
Therefore, initial singularity is absent in the past infinity.
If $c=0$, the universe begins with a flat space-time.
For cases with $n<-1$, $U(\phi) \to 0$ $(V(\phi) \to -\lambdabar)$ for large $|\phi|$.
Because the potential is almost constant, it rescales the cosmological constant to be negative, $\Lambda \to -1/\kappa< 0$.
As $|\phi|$ decreases, $U$ increases and even becomes divergent at $|\phi|=0$.
Therefore, the universe will be divided into two distinct classes according to the sign of $\phi$.
For $\phi<0$, the Hubble parameter, $H\propto -U'$, is negative.
The scale factor decreases with time from infinity and the initial state of the universe is given by a contracting phase.
On the other hand, for $\phi>0$, we have an expanding universe.
However, the early universe corresponds to the small $\phi$ area where the potential diverges at $\phi=0$.
We discard this possibility because it is not acceptable physically.
Summarizing, a nonsingular initial state exists as the exact solution~\eqref{phi:t} when the asymptotic form of the scalar potential increases not faster than the quadratic power for large field values.
Especially for $n<-1$, when the potential decreases to $-\lambdabar$ asymptotically, the initial state is given by a contracting universe which will expand after a bounce.
An extreme of this type is an exponential potential of the form $U\propto e^{\beta\phi}$.

\subsection{MPS as a fixed point in the past}
Let us examine the stability of the MPS~\eqref{phi:t} and show that it is a fixed point in the past.
We perturbs the scalar field and the Hubble parameter by adding small perturbations $\psi(t)$ and $h(t)$ as
\begin{equation}\label{1st}
\dot\phi(t) = U(\phi(t))\left[1+ \epsilon\psi(t)\right], \qquad
H(t) =-\frac{2}{3}U'(\phi(t))\left[1 + \epsilon h(t)\right],
\end{equation}
where $\phi$ and $H$ will reproduce the MPS if $\psi=0=h$.
Putting Eq.~\eqref{1st} into the equation of motions~\eqref{eom:phi} and \eqref{H} we get, to first order,
\begin{equation}\label{dpsi}
\dot \psi -2 U'(\phi) h
=0, \qquad
h = \Big(-\frac{2}{3}+\sqrt{\frac{2}{3\kappa}+\frac{8c^2}{a_0^6}}\frac{1}{U'}\Big)\psi.
\end{equation}
Combining the two equations, we have
an equation for $\psi$
$$
\frac{\dot \psi}{\psi}=- \frac43 U' +2\sqrt{\frac{2}{3\kappa}+\frac{8c^2}{a_0^6}}.
$$
Using  $\sqrt{2\lambdabar} U'=\frac{d}{dt}\log U(\phi)$ to the present order, we
get
\begin{equation} \label{psi:1}
\psi = \psi_0 U(\phi)^{-4/3}\times e^{ t/t_c} .
\end{equation}
where $t_c =1/(2\sqrt{\frac{2}{3\kappa}+\frac{8c^2}{a_0^6}})$.
Using the relation $UU' = \dot U$, we get
\begin{eqnarray}
\psi &=& \psi_0\frac{ e^{t/t_c}}{ [U(\phi)]^{4/3}}, \qquad
h = \psi_0 \left[-\frac23 +2\sqrt{\frac{2}{3\kappa}+\frac{8c^2}{a_0^6}}\frac{1}{U'(\phi)}\right] \frac{ e^{t/t_c}}{ [U(\phi)]^{4/3}}, \nn \\
\dot \beta_\pm &=& \frac{c_\pm}{a_0^3} \big(1-\epsilon \psi- 3\epsilon \int ^t h(t') dt' \big ).
\end{eqnarray}

For the specific form of potential~\eqref{U:n} and solutions~\eqref{sol:n}, the perturbation becomes,
\begin{equation}\label{pert}
U(\phi)\psi = \psi_0 \mu^{-\frac13} (n\mu t)^{\frac{n+1}{3n}} e^{\frac{t}{t_c}} .
\end{equation}
For $n>0$, the perturbations vanish at $t=0$ where $\phi \to -\infty$.
The perturbations grow with time exponentially.
For $ n < 0$, the time runs in $(-\infty,0]$.
The perturbations exponentially grow for $t< (n+1) t_c/(-3n)$ but start to decrease for $t> (n+1) t_c/(-3n)$.
For both cases, the perturbations exponentially grow at early times.
This implies that the MPS is a fixed point in the past.
For $n<-1$, the perturbations vanish at $t=0$ too, when the scale factor goes to zero.
Because the scale factor monotonically decreases, this corresponds to a later time attractor shrinking to a singularity.
Note, however, that this is because the scalar potential has unphysical form.

\section{Summary and discussion}
In this work, we study the initial state of an anisotropic universe driven by a scalar field having various potentials with its limiting form $V(\phi) \propto \phi^{2n+2}$ in EiBI gravity.
We identify the maximal pressure solution (MPS) in Eqs.~\eqref{phi:t}, \eqref{a:t} and \eqref{beta} as the initial state of the universe.
We show that the initial state of the universe is nonsingular for $n\leq 0$ and that the early universe inflates for $-1< n \leq 2$.
The anisotropy does not generate any space-time singularity, which is contrary to the case in GR.

Let us examine the equation of state of the scalar field for MPS.
The equation of state for MPS is given by $w =p/\rho= \lambdabar/(2V(\phi)+\lambdabar).$
For $n>-1$, the potential diverges at early time.
Therefore, the equation of state takes after that of the dust.
In the presence of a dust, the universe experiences a de Sitter expansion in the early universe~\cite{Cho:2012vg}.
However, in the presence of a scalar field, the de Sitter state happens only if $n=0$.
For potentials with higher or lower powers, diverse behaviors appear.
Therefore, the perfect fluid analogy is inappropriate to describe the early universe behavior.
Mathematically, this is because the Hubble parameter in Eq.~\eqref{H} is mainly dependent on the first term of Eq.~\eqref{H}, which is absent in the perfect fluid counterpart.
For MPS, the curvature scale satisfies $H\propto \phi^n$.
The curvature scale goes to infinity at initial times for $n>0$.
On the other hand, it is finite for $n\leq 0$.
Therefore, the initial state of the universe can be described with a classical gravity+quantum matter fields for $ n\leq 0$, which is a merit of EiBI gravity.

It is interesting to ask whether the present results hold or not for a universe driven by other fields such as fermions or gauge fields.
In the case of a free theory, any fermion and gauge field satisfy the Klein-Gordon equation. In addition, the {\it maximal pressure condition}~\eqref{cond}, which represents the boundary between the allowed and forbidden regions, is determined by the energy and
pressure densities only irrespective of its physical contents.
Since the present `initial state' is given by the condition, similar states will exist even for such universes.
At present, we cannot distinguish whether the initial state of our universe is regular or not because we do not know the asymptotic form of matter field which rules the early universe.
The quarks has asymptotic freedom in high density.
If a matter dominating the early universe has similar asymptotic freedom, then the initial state of the universe will be regular.
On the other hand, if the interactions between the matters enhance for high energy, then the initial state will be singular.
At the present work, we follow the original proposal of the EiBI gravity in Ref.~\cite{Banados:2010ix} where the matter is independent from the connection.
If one relax this assumption, a more general version of EiBI gravity with richer phenomenology will appear.
It would be interesting to study whether the theory may evade the initial singularity happening in the presence of fields interacting with the connection.

\section*{Acknowledgement}
This work was supported by the National Research Foundation of Korea grants funded by the Korea government NRF-2013R1A1A2006548.


\begin{thebibliography}{99}
\bibitem{Boyle:2005ug}
  L.~A.~Boyle, P.~J.~Steinhardt and N.~Turok,
  Phys.\ Rev.\ Lett.\  {\bf 96}, 111301 (2006)  [astro-ph/0507455].  

\bibitem{1983Natur.304...39P}
D.~N. Page, Nature {\bf 304}, 39 (1983).

\bibitem{Carroll:2005it}
  S.~M.~Carroll and J.~Chen,
  Gen.\ Rel.\ Grav.\  {\bf 37}, 1671 (2005)  [Int.\ J.\ Mod.\ Phys.\ D {\bf 14}, 2335 (2005)]  [gr-qc/0505037].

\bibitem{Vilenkin}
A. Vilenkin, Phys. Rev. D {\bf 27}, 2848 (1983).

\bibitem{HH}
J. Hartle and S. Hawking,
Phys. Rev. D {\bf 28}, 2960 (1983).

\bibitem{Banados:2010ix}
  M.~Banados and P.~G.~Ferreira,
  Phys.\ Rev.\ Lett.\  {\bf 105}, 011101 (2010)
  [arXiv:1006.1769 [astro-ph.CO]].

\bibitem{Pani:2011mg}
  P.~Pani, V.~Cardoso and T.~Delsate,
  Phys.\ Rev.\ Lett.\  {\bf 107}, 031101 (2011)
  [arXiv:1106.3569 [gr-qc]].

\bibitem{Pani:2012qb}
  P.~Pani, T.~Delsate and V.~Cardoso,
  Phys.\ Rev.\ D {\bf 85}, 084020 (2012)
  [arXiv:1201.2814 [gr-qc]].

\bibitem{DeFelice:2012hq}
  A.~De Felice, B.~Gumjudpai and S.~Jhingan,
  Phys.\ Rev.\ D {\bf 86}, 043525 (2012)
  [arXiv:1205.1168 [gr-qc]].

\bibitem{Avelino:2012ge}
  P.~P.~Avelino,
  Phys.\ Rev.\ D {\bf 85}, 104053 (2012)  [arXiv:1201.2544 [astro-ph.CO]].


\bibitem{Avelino:2012qe}
  P.~P.~Avelino,
  JCAP {\bf 1211}, 022 (2012)
  [arXiv:1207.4730 [astro-ph.CO]].

\bibitem{Casanellas:2011kf}
  J.~Casanellas, P.~Pani, I.~Lopes and V.~Cardoso,
  Astrophys.\ J.\  {\bf 745}, 15 (2012)
  [arXiv:1109.0249 [astro-ph.SR]].

\bibitem{EscamillaRivera:2012vz}
  C.~Escamilla-Rivera, M.~Banados and P.~G.~Ferreira,
  Phys.\ Rev.\ D {\bf 85}, 087302 (2012)
  [arXiv:1204.1691 [gr-qc]].

\bibitem{Avelino:2012ue}
  P.~P.~Avelino and R.~Z.~Ferreira,
  Phys.\ Rev.\ D {\bf 86}, 041501 (2012)
  [arXiv:1205.6676 [astro-ph.CO]].

\bibitem{Liu:2012rc}
  Y.~-X.~Liu, K.~Yang, H.~Guo and Y.~Zhong,
  Phys.\ Rev.\ D {\bf 85}, 124053 (2012)
  [arXiv:1203.2349 [hep-th]].

\bibitem{Delsate:2012ky}
  T.~Delsate and J.~Steinhoff,
  Phys.\ Rev.\ Lett.\  {\bf 109}, 021101 (2012)
  [arXiv:1201.4989 [gr-qc]].

\bibitem{Pani:2012qd}
  P.~Pani and T.~P.~Sotiriou,
  Phys.\ Rev.\ Lett.\  {\bf 109}, 251102 (2012)  [arXiv:1209.2972 [gr-qc]].  

\bibitem{Cho:2013pea}
  I.~Cho, H.-C.~Kim and T.~Moon,
  Phys.\  Rev.\  Lett.\ {\bf 111}, 071301 (2013)
  [arXiv:1305.2020 [gr-qc]].

\bibitem{Sham:2012qi}
  Y.~-H.~Sham, L.~-M.~Lin and P.~T.~Leung,
  Phys.\ Rev.\ D {\bf 86}, 064015 (2012)
  [arXiv:1208.1314 [gr-qc]].

\bibitem{Harko:2013wka}
  T.~Harko, F.~S.~N.~Lobo, M.~K.~Mak and S.~V.~Sushkov,
  Phys.\ Rev.\ D {\bf 88}, 044032 (2013)
  [arXiv:1305.6770 [gr-qc]].

\bibitem{Sham:2013sya}
  Y.~-H.~Sham, P.~-T.~Leung and L.~-M.~Lin,
  Phys.\ Rev.\ D {\bf 87}, 061503 (2013)
  [arXiv:1304.0550 [gr-qc]].

\bibitem{Yang:2013hsa}
  K.~Yang, X.~-L.~Du and Y.~-X.~Liu,
  arXiv:1307.2969 [gr-qc].

\bibitem{Ade:2013ktc}
  P.~A.~R.~Ade {\it et al.}  [Planck Collaboration],
  arXiv:1303.5062 [astro-ph.CO].

\bibitem{Cho:2012vg}
  I.~Cho, H.-C.~Kim and T.~Moon,
  Phys.\ Rev.\ D {\bf 86}, 084018 (2012)
  [arXiv:1208.2146 [gr-qc]].

\bibitem{Scargill:2012kg}
  J.~H.~C.~Scargill, M.~Banados and P.~G.~Ferreira,
Phys.\ Rev.\ D {\bf 86}, 103533 (2012)  [arXiv:1210.1521 [astro-ph.CO]].  

\bibitem{Cho:2013usa}
  I.~Cho and H.-C.~Kim,
  Phys.\ Rev.\ D {\bf 88}, 064038 (2013)
  [arXiv:1302.3341 [gr-qc]].
\end{thebibliography}
\end{document}